\documentstyle[12pt,epsf]{article}

\begin{document}
\title{\bf Way to Discriminate between Mesons \\ 
and Glueballs for $I=0, J^{PC}=even^{++}$ \\ 
Unflavored Hadrons}
\author{{\normalsize Chong-Shou Gao} \vspace*{6pt} \\
{\normalsize Department of Physics, Peking University, Beijing 100871,
China} \\ 
{\normalsize Institute of Theoretical Physics, Academia Sinica, Beijing
100080, China}\\
}
\maketitle
\begin{abstract}
Based on the general analysis of branching ratio of two pseudoscalar
meson channels, discriminants between mesons and glueballs for
$I=0, J^{PC}=even^{++}$ unflavored hadrons with mass between
1.2 GeV and 2.9 GeV are suggested. Known $I=0, J^{PC}=2^{++},
f_2(1525)$ particle 
is discriminated as a typical meson. The way to discriminate new
$I=0, J^{PC}=even^{++}$ unflavored hadrons is discussed.
\vspace*{8pt} \\
PACS number(s): 12.39.Mk 14.40.Cs 13.25.Jx
\end{abstract}
\quad Any $I=0, J^{PC}=even^{++}$ unflavored hadron $X$ 
can be a meson composed of light quark and
anti-quark, a hybrid composed of light quark and anti-quark and gluon, 
a two gluon glueball or a meson composed of heavy quark and anti-quark. 
If the mass of $X$ is lighter than 2.9 GeV, the possibility of being a heavy 
quarkonium should be ruled out. One may discriminate these possibilities 
with each other by
the decay behavior of this hadron. In general $X$ can decay into two
pseudoscalar mesons and the possible decay modes are: $\pi ^+ \pi^-,
\pi ^0 \pi ^-, K^+K^-, K_SK_S, K_LK_L, \eta \eta, \eta \eta ^{\prime},
\eta ^{\prime} \eta ^{\prime}$. The effective Hamiltonian of such kind of
decay modes can be described generally as \cite{1}
\begin{equation}
H_{eff}=\frac {f}{m_X^{J-!}}X_{\mu \nu ...\lambda \sigma}(\partial _1^\mu -\partial _2^\mu
)(\partial _1^\nu -\partial _2^\nu )...(\partial _1^\lambda -\partial
_2^\lambda )(\partial _1^\sigma -\partial _2^\sigma )P_1P_2,
\end{equation}
where $P_1$ and $P_2$ describe peudoscalar mesons in final state and
$X_{\mu \nu ...\lambda \sigma}$ with J subscripts describes the
particle $X$ of spin $J$ and satisfies constraints: $X_{\mu \nu
...\lambda \sigma}$ is fully symmetric in $\nu \mu 
...\lambda \sigma$; $g^{\mu \nu}X_{\mu \nu ...\lambda \sigma}=0$ and
 $\partial ^\mu X_{\mu \nu ...\lambda \sigma}=0$.

The partial width $\Gamma (P_1P_2)$ of decay
mode $X \rightarrow P_1+P_2$ can be expressed as
\begin{equation}
\Gamma (P_1P_2) = \eta \alpha \frac{k^{2J+1}}{m_X^{2J}},
\end{equation}
where
$k$ is the decay momentum in the center of mass system, $m_X$ is the
mass of $X$ particle, $\alpha $ is the dimensionless effective coupling
constant
\begin{equation}
\alpha = \frac {f^2}{4\pi }
\end{equation}
$\eta $ is a numerical coefficient which takes the value of
\begin{equation}
\eta = \frac{8^J(J!)^2}{2(2J+1)!},
\end{equation}

If $X$ is a meson composed of light quark and anti-quark, the flavor
structure of effective interaction can be described generally as
\begin{equation}
\begin{array} {ll}
H_{eff}=&\frac {1}{m_X^{J-1}}[f_1Tr(XPP)+f_2Tr(X)Tr(PP)\\
&+f_3Tr(XP)Tr(P)+f_4Tr(X)Tr(P)Tr(P)],
\end{array}
\end{equation}
where $X$ and $P$ are $3\times 3$ flavor matrices which have the form
\begin{equation}
X=X_J\left ( \begin{array} {ccc}
1&0&0\\
0&1&0\\
0&0&A
\end{array}\right ).
\end{equation}
where
$A=1$ or $A=-2$ means $X$ is a singlet or an octet of flavor $SU(3)$
group respectively, $A$ takes other values means $X$ belongs to a
mixing state $\bf 1+\bf 8$ of flavor $SU(3)$ group.
\begin{equation}
P=\left( \begin{array}{ccc}
\frac{\pi ^0}{\sqrt{2}}+\alpha \eta +\beta \eta '\quad &\pi ^+\quad &K^+\\
\pi ^-\quad &-\frac{\pi ^0}{\sqrt{2}}+\alpha \eta +\beta \eta '\quad &K^0\\
K^-\quad &\overline {K}^0\quad &-\sqrt{2}\beta \eta +\sqrt{2}\alpha \eta '\\
\end{array}\right) .
\end{equation}
where
\begin{equation}
\alpha =\frac{\cos \theta - \sqrt{2} \sin \theta}{\sqrt{6}},\\
\beta =\frac{\sin \theta + \sqrt{2} \cos \theta}{\sqrt{6}}.
\end{equation}
$\theta$ is the mixing angle of $\bf 1+\bf 8$ for pseudoscalar mesons.

The values of coupling constants $f_1$, $f_2$, $f_3$ and $f_4$ depend
on the dynamical mechanism of specific decay modes, and are independent to
each other. The first term of the effective interaction involving $f_1$
describes the Zweig allowed  decay while the other terms  
involving $f_2$, $f_3$ and $f_4$ describe the Zweig
suppressed terms. Thus $f_2$, $f_3$ and $f_4$ are expected to be much smaller
than $f_1$ and their contributions to the decay width should be negligible
in comparison with $f_1$. The flavor structure of the effective interaction
of $X$ meson decaying into two pseudoscalar mesons can be described
simply as
\begin{equation}
H_{eff}=\frac {f_1}{m_X^{J-1}}Tr(XPP),
\end{equation}
The dimensionless effective coupling constant $\alpha _{eff}$ can be
expressed as
\begin{equation}
\alpha _{eff}=G(P_1P_2)\alpha _1,
\end{equation}
where the coefficient $G(P_1P_2)$ can be calculated from above
effective interaction as in Table 1.

\begin{center}
Table 1 \quad Coefficient $G(P_1P_2)$ of decay
mode $X \rightarrow P_1+P_2$ for meson $X$.
\vspace*{4pt}

\begin{tabular} {|c|c|c|c|}
\hline
Modes & $G(P_1P_2)$ & Modes & $G(P_1P_2)$ \\ \hline
$\pi ^0\pi ^0$ & $2$ & $K^+K^-$ & $(1+A)^2$ \\ \hline
$\pi ^+\pi ^-$ & $4$ & $\eta \eta$ & $8(\alpha ^2+A\beta ^2)^2$ \\ \hline
$K_SK_S$ & $\frac{(1+A)^2}{2}$ & $\eta \eta '$ & $ 16\alpha ^2\beta
^2(1-A)^2$ \\ \hline
$K_LK_L$ & $\frac{(1+A)^2}{2}$  &$\eta '\eta '$ & $8(\beta ^2+A\alpha ^2)^2$
\\ \hline
\end{tabular}
\end{center}

If one introduces $R(\frac{P_1P_2}{P_3P_4})$
\begin{equation}
R(\frac {P_1P_2}{P_3P_4})=\frac{\alpha _{eff}(P_1P_2)}{\alpha
_{eff}(P_3P_4)} ,
\end{equation}
two discriminants of $X$ being a pure meson composed by light quark and anti-quark 
can be obtained as
\begin{equation}
R(\frac {\eta \eta}{K^+K^-})=\frac{2(\alpha ^2-\beta
^2)^2}{R(\frac{K^+K^-}{\pi ^+\pi ^-})}+\frac{8(\alpha ^2-\beta ^2)\beta
^2}{\sqrt{R(\frac{K^+K^-}{\pi ^+\pi ^-})}}+8\beta ^4,
\end{equation}
and
\begin{equation}
R(\frac {\eta \eta}{K^+K^-})+R(\frac {\eta \eta '}{K^+K^-})+R(\frac
{\eta '\eta '}{K^+K^-})=2+\frac{1}{R(\frac{K^+K^-}{\pi ^+\pi
^-})}-\frac{2}{\sqrt{R(\frac{K^+K^-}{\pi ^+\pi ^-})}}.
\end{equation}

These two discriminants are independent of whether $X$ ia a pure
singlet, belongs to a octet or is a mixing state of $\bf 1+\bf 8$
representation of flavor $SU(3)$ group.

>From these relations one may find that if $R(\frac{K^+K^-}{\pi ^+\pi
^-})\ge 0.25$, then $R(\frac {\eta \eta}{K^+K^-})+R(\frac {\eta \eta
'}{K^+K^-})+R(\frac
{\eta '\eta '}{K^+K^-})\le 2.0$. So, if experimental data shows that $R(\frac
{\eta \eta}{K^+K^-
})+R(\frac {\eta \eta '}{K^+K^-})+R(\frac
{\eta '\eta '}{K^+K^-})>>2.0$, the possibility of $X$ being a pure meson
state should be ruled out.

If $X$ is a hybrid composed of light quark, anti-quark and a gluon, the
flavor structure of effective interaction can be described in 
the same way as that of the meson case, so it is difficult to distinguish
hybrid from meson by its behavior of these decay modes.

If $X$ is a glueball composed of two gluons, the flavor
structure of effective interaction can be described generally as
\begin{equation}
H_{eff}=\frac {1}{m_X^{J-1}}[f_2X_JTr(PP)+f_4X_JTr(P)Tr(P)],
\end{equation}

The values of coupling constants $f_2$ and $f_4$ depend
on the dynamical mechanism of specific decay modes, and are independent to
each other. Both terms of the effective interaction will give contribution
to specific decay modes.

If the term involving $f_2$ dominates the decay, the coefficient
$G(P_1P_2)$ can be calculated from above
effective interaction and the results are listed in Table 2.

\begin{center}
Table 2 \quad Coefficient $G(P_1P_2)$ of decay mode $X \rightarrow P_1+P_2$\\
 for glueball $X$ with $f_2$ term dominates.
\vspace*{4pt}

\begin{tabular} {|c|c|c|c|}
\hline
Modes & $G(P_1P_2)$ & Modes & $G(P_1P_2)$ \\ \hline
$\pi ^0\pi ^0$ & $2$ & $K^+K^-$ & $4$ \\ \hline
$\pi ^+\pi ^-$ & $4$ & $\eta \eta$ & $2$ \\ \hline
$K_SK_S$ & $2$ & $\eta \eta '$ & $0$ \\ \hline
$K_LK_L$ & $2$  &$\eta '\eta '$ & $2$ \\ \hline
\end{tabular}
\end{center}

Four discriminants of $X$ being a glueball with $f_2$ term dominate 
can be obtained as
\begin{equation}
R(\frac {K^+K^-}{\pi ^+\pi ^-})=1,\\
\end{equation}
\begin{equation}
R(\frac {\eta \eta }{K^+K^-})=\frac {1}{2},\\
\end{equation}
\begin{equation}
R(\frac {\eta \eta '}{K^+K^-})=0,\\
\end{equation}
and
\begin{equation}
R(\frac {\eta '\eta '}{K^+K^-})=\frac {1}{2}.
\end{equation}

If the term involving $f_4$ dominates the decay, the coefficient
$G(P_1P_2)$ can be calculated from the above
effective interaction as shown in Table 3.

\begin{center}
Table 3 \quad Coefficient $G(P_1P_2)$ of decay mode $X \rightarrow P_1+P_2$\\
  for glueball $X$ with $f_4$ term dominates.
\vspace*{4pt}

\begin{tabular} {|c|c|c|c|}
\hline
Modes & $G(P_1P_2)$ & Modes & $G(P_1P_2)$ \\ \hline
$\pi ^0\pi ^0$ & $0$ & $K^+K^-$ & $0$ \\ \hline
$\pi ^+\pi ^-$ & $0$ & $\eta \eta$ & $ 18\sin ^4\theta $ \\ \hline
$K_SK_S$ & $0$ & $\eta \eta '$ & $ 36\sin ^2\theta \cos ^2\theta $ \\ \hline
$K_LK_L$ & $0$  &$\eta '\eta '$ & $ 18\cos ^4\theta $ \\ \hline
\end{tabular}
\end{center}

This term leads to only three decay modes: $\eta \eta $,  $\eta \eta '$
and  $\eta '\eta '$. In general,  both terms of interaction will give
contributions to the decay, the effective coupling constant should be
the sum of these two terms. Then, in general, three discriminants of
glueball should be changed as:
\begin{equation}
R(\frac {K^+K^-}{\pi ^+\pi ^-})=1,\\
\end{equation}
\begin{equation}
R(\frac {\eta \eta }{K^+K^-})\ge \frac {1}{2},\\
\end{equation}
and
\begin{equation}
R(\frac {\eta \eta}{K^+K^-})+R(\frac {\eta \eta '}{K^+K^-})+R(\frac
{\eta '\eta '}{K^+K^-})\ge 1.
\end{equation}

$R(\frac {P_1P_2}{P_3P_4})$ canbe obtained from partial widths of these
two decay modes by
\begin{equation}
R(\frac {P_1P_2}{P_3P_4})=\frac {\Gamma
(P_1P_2)k_{P_3P_4}^{2J+1}}{\Gamma (P_3P_4)k_{P_1P_2}^{2J+1}},
\end{equation}
and $\alpha $ and $\beta $ are known functions of the mixing angle
$\theta $ of pseudoscalar mesons, thus all quantities in (12), (13), (19), 
(20) and (21) can be obtained directly from experiments. These five 
relations can be treated as a set of discriminants for whether a $I=0,
J^{PC}=even^{++}$ unflavored hadrons with mass between 1.2 GeV and 2.9
GeV is a meson, a hybrid or a two gluon glueball.

It is important to note that for the case of $R(\frac{K^+K^-}{\pi ^+\pi
^-}) \ne 1.00$, the possibility of $X$ being a pure glueball should be ruled 
out. But if $X$ is a pure light quarkonium or a hybrid with light quark and 
antioquark, the values of $R(\frac {\eta \eta}{K^+K^-})$ and
$R(\frac {\eta \eta}{K^+K^-})+R(\frac {\eta \eta '}{K^+K^-})+R(\frac
{\eta '\eta '}{K^+K^-})$ should satisfy equations (12) and (13) respectively. 
If the experimental values of $R(\frac {\eta \eta}{K^+K^-})$ and
$R(\frac {\eta \eta}{K^+K^-})+R(\frac {\eta \eta '}{K^+K^-})+R(\frac
{\eta '\eta '}{K^+K^-})$ are
larger than the corresponding values expected from (12) and (13) for the 
pure meson or hybrid case, $X$ should be a mixing state of glueball and 
light quarkonium or hybrid. Thus (12), (13), (19), (20) and (21)  can be adopted to
judge whether a particle is a pure glueball, 
a pure meson or hybrid, or a mixing state of glueball and meson or hybrid 
unambiguously.

\begin{figure}
\epsfxsize=16cm
\centerline{\epsffile{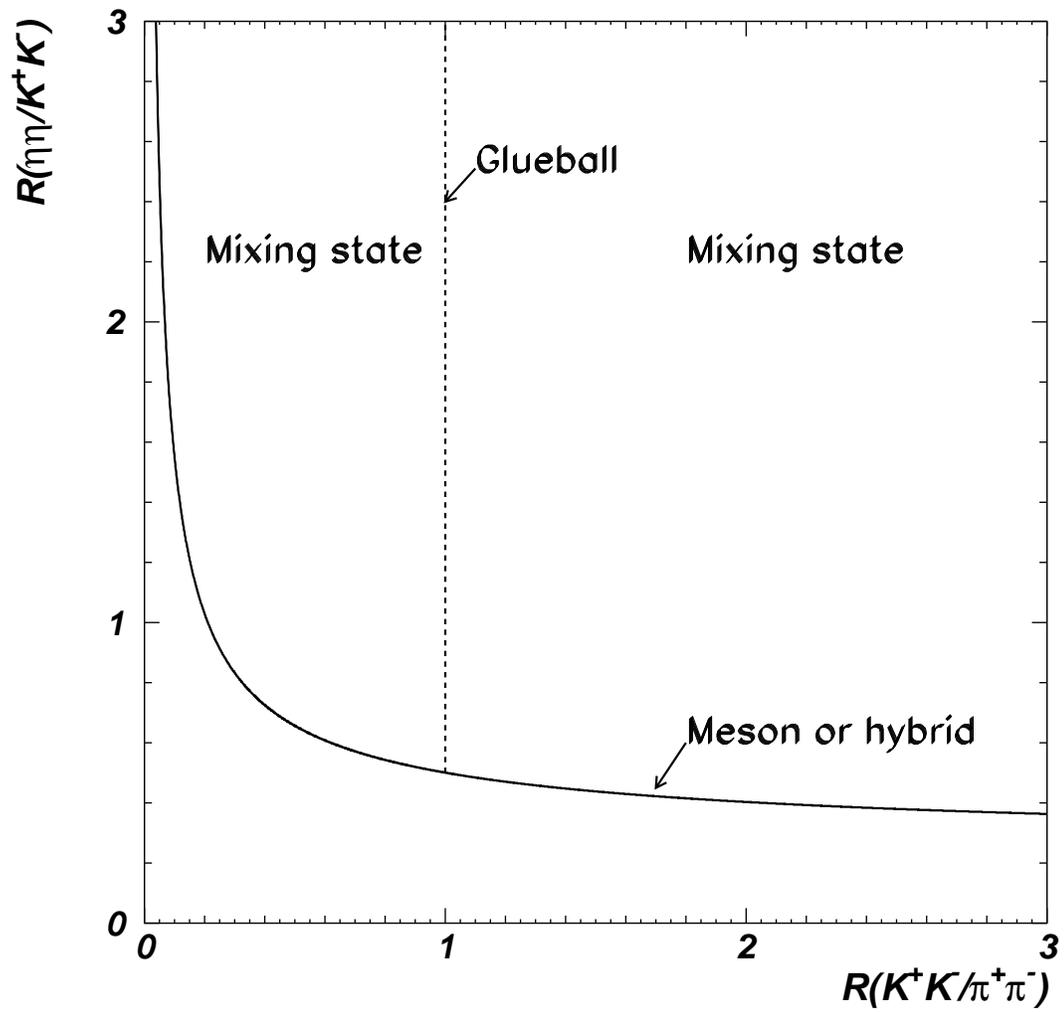}}
\caption[]{Discriminant via $R(\frac{\eta \eta}{K^+K^-})$
and $R(\frac{K^+ K^-}{\pi^+ \pi^-})$.}
\end{figure}

\begin{figure}
\epsfxsize=16cm
\centerline{\epsffile{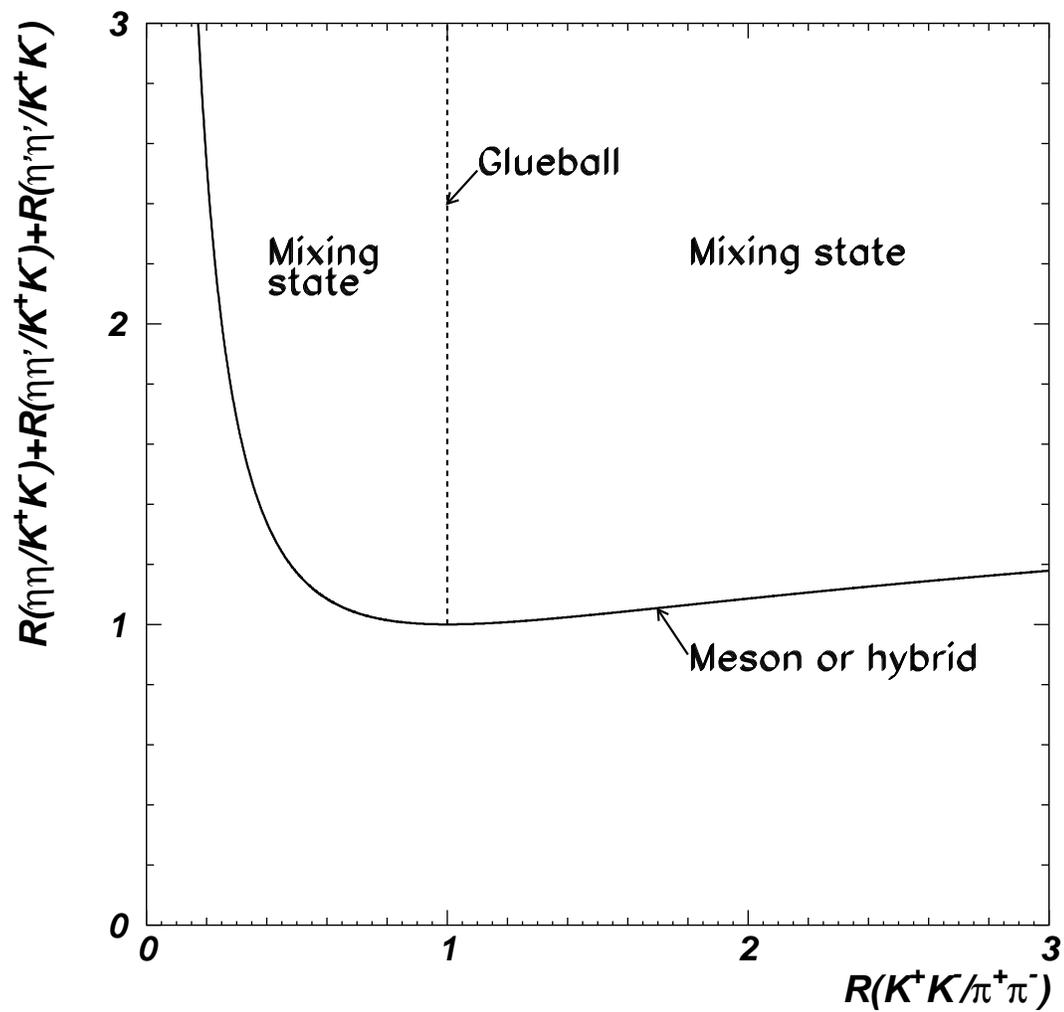}}
\caption[]{Discriminant via $R(\frac{\eta \eta}{K^+K^-})
+ R(\frac{\eta \eta '}{K^+K^-})
+ R(\frac{\eta ' \eta '}{K^+K^-})$
and $R(\frac{K^+ K^-}{\pi^+ \pi^-})$.}
\end{figure}

Above discriminants can be described in Fig 1 and Fig 2. If the values of $R(\frac {\eta \eta}{K^+K^-})$ and
$R(\frac {\eta \eta}{K^+K^-})+R(\frac {\eta \eta '}{K^+K^-})+R(\frac
{\eta '\eta '}{K^+K^-})$ are laid on the solid lines in 
Fig 1 and Fig 2, $X$ should be a pure light quarkonium or a haybrid.  
If the values of $R(\frac {\eta \eta}{K^+K^-})$ and
$R(\frac {\eta \eta}{K^+K^-})+R(\frac {\eta \eta '}{K^+K^-})+R(\frac
{\eta '\eta '}{K^+K^-})$ are laid on the dashed lines in 
Fig 1 and Fig 2, $X$ should be a pure glueball.  If the values of $R(\frac {\eta \eta}{K^+K^-})$ and
$R(\frac {\eta \eta}{K^+K^-})+R(\frac {\eta \eta '}{K^+K^-})+R(\frac
{\eta '\eta '}{K^+K^-})$ are laid above the solid lines and not on the dashed lines in 
Fig 1 and Fig 2, $X$ should be a mixing state of light quarkonium or haybrid and glueball.  

One well-known particle $f_2(1525)$ is an $I=0, J^{PC}=2^{++}$
unflavored hadron. Its branching ratios of two pseudoscalar decay modes
are $Br(K\overline K)=(88.8\pm 3.1)\%$, $Br(\eta \eta)=(10.3\pm 3.1)\%$ and
$(\pi \pi)=(0.82\pm 0.15)\%$, respectively \cite{2}. Thus one gets $R(\frac
{K^+K^-}{\pi _+\pi _-})=(290\pm 38)$. According to the above discriminants,
the possibility of $f_2(1525)$ being a pure glueball should be ruled out. If $R(\frac {\eta \eta}{K^+K^-})=0.224$, $f_2(1525)$ should be a pure meson
or a hybrid; if $R(\frac {\eta \eta}{K^+K^-})\gg 0.224$, $f_2(1525)$
should be a mixing state. From above data one gets $R(\frac {\eta
\eta}{K^+K^-})=(0.36\pm 0.11)$, so it leads to the conclusion that
$f_2(1525)$ is a pure meson or hybrid.

One new unflavored particle $f_J(2220)$ was observed in radiative decay
of $J/\psi $ meson \cite{3}. The spin of  $f_J(2220)$ is $J = 2$ or $4$. The combined branching ratios of two
pseudoscalar decay modes are \cite{4,5}
\begin{equation}
Br(J/\psi \rightarrow \gamma f_J(2220))Br(f_J(2220) \rightarrow \pi
^+\pi ^-) = (5.6_{-1.6}^{+1.8}\pm 2.0)\times 10^{-5},
\end{equation}
\begin{equation}
Br(J/\psi \rightarrow \gamma f_J(2220))Br(f_J(2220) \rightarrow K^+K^-)
= (3.3_{-1.3}^{+1.6}\pm 1.2)\times 10^{-5}, 
\end{equation}
\begin{equation}
Br(J/\psi \rightarrow \gamma f_J(2220))Br(f_J(2220) \rightarrow
K^0_sK^0_s ) = (2.7_{-0.9}^{+1.1}\pm 0.8)\times 10^{-5},
\end{equation}

This means that if the spin of  $f_J(2220)$ is $J = 2$, then
\begin{equation}
R(\frac{K^+K^-}{\pi ^+\pi ^-}) = 0.98 \pm 0.72.
\end{equation}
>From (12) and (13) one may get discriminants that if $f_J(2220)$ is a pure 
meson or hybrid, the values of $R(\frac {\eta \eta}{K^+K^-})$ and
$R(\frac {\eta \eta}{K^+K^-})+R(\frac {\eta \eta '}{K^+K^-})+R(\frac
{\eta '\eta '}{K^+K^-})$ should be 
\begin{equation}
R(\frac {\eta \eta}{K^+K^-}) = 0.50_{-0.06}^{+0.40} 
\end{equation}
and
\begin{equation}
R(\frac {\eta \eta}{K^+K^-})+R(\frac {\eta \eta '}{K^+K^-})+R(\frac
{\eta '\eta '}{K^+K^-}) =1.00_{-0.00}^{+0.95}.
\end{equation}
>From (19), (20) and (21) one may get discriminants that if $f_J(2220)$ is a pure 
glueball the values of $R(\frac {\eta \eta}{K^+K^-})$ and
$R(\frac {\eta \eta}{K^+K^-})+R(\frac {\eta \eta '}{K^+K^-})+R(\frac
{\eta '\eta '}{K^+K^-})$ should be 
\begin{equation}
R(\frac {\eta \eta}{K^+K^-}) \gg 1.70 
\end{equation}
and
\begin{equation}
R(\frac {\eta \eta}{K^+K^-})+R(\frac {\eta \eta '}{K^+K^-})+R(\frac
{\eta '\eta '}{K^+K^-}) \gg 3.85.
\end{equation}
Thus it is important to investigate the $\eta \eta $, $\eta \eta '$ and $\eta '\eta '$ 
decay modes of $f_J(2220)$ particle.  

In summary, four discriminants based on the observed branching ratios of
two pseudoscalar meson decay modes for a $I=0, J^{PC}=even^{++}$
unflavored hadron $X$ with mass between 1.2 GeV and 2.9 GeV can be used
to judge whether $X$ particle is a pure meson or hybrid, a pure glueball or a 
mixing state.

\section*{Acknowledgments}

\qquad I would like to thank Prof. X. D. Ji, Prof. T. Cohen, 
Dr. Y. N. Gao, Dr. H. B.
Hu, Dr. S. H. Zhu and Mr. K. Z. Gao for helpful discussions and
suggestions. This 
work was supported in part by the National Natural Science Foundation
of China and Doctoral Program Foundation of Institution of Higher Education.

\end{document}